# Functional Imaging for Dose Painting in Radiotherapy

Yaru Pang[1], Gary Royle[1], Spyros Manolopoulos[1]

[1] Department of Medical Physics and Biomedical Engineering, University College London, London, WC1E 6BT, United Kingdom

## Abstract:

Dose painting has been developed to modulate the required dose in target area without increasing the toxicity in healthy areas. Apart from determining the accurate location and size of tumours, quantitative functional imaging can be used to implement the dose painting. Functional imaging, such as multi parameter MRI and PET CT, allows us to achieve biological dose escalation by increasing the dose in certain areas or voxels that are therapy-resistant in the gross tumor volume, while reducing the dose in less aggressive area or voxels. Functional imaging can serve as an indicator of therapeutic intervention in radiotherapy due to microscopic tissue properties. With such biological indicators, personalized radiation dose can be tailored to a specific contour or a voxel using dose painting. In this review, we firstly discuss several quantitative functional imaging techniques including PET-CT and multi-parameter MRI. Furthermore, theoretical and experimental comparisons for dose painting by contours (DPBC) and dose painting by numbers (DPBN), along with outcome analysis after dose painting are provided. Finally, we conclude major challenges and future directions in this field though which we hope to inspire exciting developments and fruitful research avenues.

## I. Introduction

For treatment planning and outcome management, medical imaging is the key to guarantee the best outcome of radiation therapy (RT) [1]. However, the majority of imaging approaches such as CT and MRI only determine tumour size and location. Conventional imaging techniques have limitations to achieve deep insight into tumour marco- and micro-environments. To acquire such information for evaluating the severity of diseases, quantitative functional imaging has been put forward by extracting their quantifiable radiologic biomarkers [2]. Moreover, the role of functional images used in protecting function of critical tissue and neural nerves when performing RT has been explored, which leads to a thriving research and development direction.

RT continues to be a standard treatment for malignant tumors, resulting in improved outcomes over surgery and chemotherapy alone. Moreover, RT technology has been developed rapidly over the past decades. However, local recurrence after RT is one of the important modes of failure of most malignant tumors. The main reason may be that the planning target volume (PTV) of the primary tumor receives a uniform prescribed dose without considering the heterogeneity of the tumor itself in terms of time and space. Therefore, dose painting was originally proposed at ESTRO in 1998. In 2000, Ling et al. employed biological imaging to achieve "biological conformality", where higher doses are applied to some parts of the tumor area with higher clonogenic cell density and radiation resistance, while lower doses to less aggressive parts [6,7]. Therefore, tumor cells can be eliminated, and healthy tissues can be recovered [3,4]. To improve the accuracy of dose painting, many functional imaging approaches have been proposed recently [2]. Functional images have potential strengths to improve prognostication response to RT, which can facilitate personalized treatment and clinical trial designs in terms of patient-specific prescription dose and biological target volume (BTV) [6]. Moreover, quantitative functional imaging can be used for heterogeneous dose painting, where doses can be spatially redistributed throughout the target tumour based on personalized parameter maps [1].

Although current quantitative-imaging techniques are largely used for response management, there are only very limited studies on dose painting and no prior studies related to dose stratification. Therefore, clinical potential of quantitative imaging is a prosperous area that deserves investigations [1]. Heide et al. suggested that high-quality imaging of the tumor and its surrounding tissue, facilitates effective dose painting [2]. They discussed MRI-guided dose painting in 2012. With the development of the functional imaging, it can become a promising index to predict the dose painting. In this article, we summarize various functional images as the pre-requisite for dose painting, as shown in table 1. The details of each image modality are discussed in Section 2. After that, dose painting by contours (DPBC) and dose painting by numbers (DPBN) using these functional imagings are discussed in Section 3. Although at present, the supplementary exposure of dose painting to biological target areas is mostly implemented through PET-CT imaging technology, other modalities still have potential in dose painting.

Table 1. An overview of functional imaging techniques

| Functional imaging techniques | Quantitative parameters | Biomarkers | Threshold |
|---|---|---|---|
| MR-Spectroscopic (MRS) | Metabolism | A ratio of choline to NAA (Cho/NAA) | Not clear |
| Diffusion weighted (DW)-MRI | Diffusion of water molecules | Apparent diffusion coefficient (ADC) | Not clear |
| Perfusion MRI | Tissue perfusion | Cerebral blood volume (CBV), cerebral blood flow (CBF), transfer constant of Gd-diethylenetriamine pentaacetic acid (Ktrans) | relative (r)CBV > 1.75 |
| Diffusion tensor imaging (DTI)- MRI | Tensor of water diffusion | White matter tracts (WMT) | Not clear |
| PET-CT | Glucose metabolism and the upregulation of glucose transporters in cancer cells | Standardized uptake value (SUV) | Not clear |

## II. Functional Imaging

Since there are many imaging techniques used for dose painting, the widely discussed are DW-MRI, MRS, Perfusion MRI including Dynamic susceptibility contrast (DSC) and dynamic contrast enhanced (DCE), DTI MRI and PET-CT. We present a review for these methods by surveying the state-of-the-art works.

### 2.1 DW-MRI

Diffusion-weighted magnetic resonance imaging (DW-MRI) is used to measure the mobility of water molecules in the microscopic environment of tissues. DW-MRI is very sensitive to cellular density, proliferation power and cellular permeability [8,9], therefore this technique can reveal microscopic details of normal and diseased tissues. The sensitized signal is modelled by the amount of diffusion weights, called b-value. The amount of diffusion existing in the tissue, predominantly in the extracellular space [10, 11], is known as the apparent diffusion coefficient (ADC) [12]. ADC map is an MRI image that shows better diffusion than conventional DWI. ADC map are achieved by DW-MRI with at least two b-values [13]. Versus DWI images, darker areas in the ADC images represents smaller magnitudes of diffusion. Moreover, the lower ADC value indicates the slower water infusing where malignant tumour appears [14]. Therefore, ADC and cellular density have an inverse correlation relationship. Several studies estimated the relationship between ADC and cellular density for different site of tumour. Ginat et al. performed a histological study and achieved the relationship between ADC and cellular density for chordoma [15], while Gupta et al. estimated the relationship for GBM [16].

The restricted spread of water has been considered an indicator of solid tumours. However, in DW-MRI with b values of 0 and 800-1000s/mm$^2$, the ADC values are usually unpredictable because of the mixture of micronecrosis, normal tissues, high-cell tumors and edema. The ADC values of tumor are sometimes even higher than those in normal tissues. Therefore, researchers used the minimum ADC value to determine its prognostic value [17-19]. However, the area of malignant tumours is usually larger than the area with the minimum ADC value. Therefore, Mardor et al. used high b-value DW-MRI (e.g. 3000-4000s/mm$^2$) to eliminate the signal from edema [21-24]. Pramanik et al. showed that the hypercellular subvolume (HCV) of GBM determined by high-b value (3000s/mm$^2$) DWI can predict progression-free survival (PFS) [24]. About 40% of HCV may exceed the area of the traditional high-dose volume, which suggests that it may be a potential biological target with more radiation dose [24]. Figure 1 is an example of DW-MRI with two different b-values. There are two enhanced lesions in a GBM 19 months post-RT.

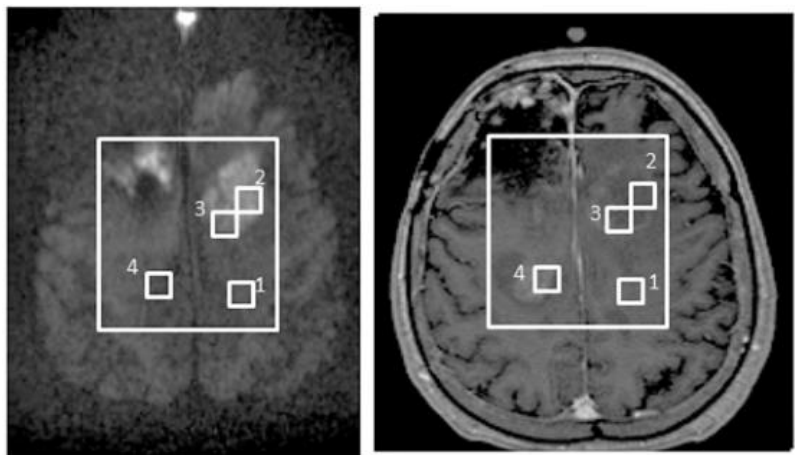

Figure 1. DW-MRI with two different b-values [30].

The left figure is DW-MRI with high b-value (b=3000s/mm$^2$) while the right one is 800-1000s/mm$^2$. The two new enhanced lesions in boxes 2, 3 were more obvious in the left figure and has been proved by MR-Spectroscopic Imaging (MRS) [30].

Hamstra et al. [25] and Moffat et al. [26] provided information for individual patients' adaptation to radiotherapy. Patients usually have higher overall survival (OS), whose voxel ADC values are changed dramatically after 3 weeks of radiotherapy. However, when the tumor grows or shrinks, the paired images obtained before and during radiotherapy must consistently produce high-quality registration. [27]. ADC can also be used to calculate tumour control probability (TCP) to analyse patient-specific characteristics. A recent study [28] showed that MRI-driven cellular density can enhance TCP value differences in patients. Buizza et al. evaluated DW-MRI for modelling TCP in skull-base chordomas, which has enrolled in CIRT protocol [29]. The aforementioned methods facilitate personalized and optimized treatments. The main limitation of these studies lies in uncertainties that inherently affect the relationship between ADC and cellular density. Single-shot echo planar imaging (EPI) pulse sequence is used in DW-MRI, however, it is very sensitive to geometric distortion. To reduce the geometric distortion due to EPI, multi-shot EPI, read-out segmented multi-shot EPI, and high parallel imaging factor are used. To achieve the golden standard, tumor target defined by high b-value DWI and parameter response curves therefore requires pathological verification.

## 2.2 MRS

MR-Spectroscopic Imaging (MRS) uses radiolabeled glucose and methionine to identify high-risk regions in large tumours. In MRS, active tumors exhibit at the areas with high resonance in the choline spectral peak and a low NAA (N-acetylaspartate). In other words, creatine resonance correlating with high choline/NAA, or choline/creatine ratios versus low ratios for areas of inactivity [31, 32-38].

Graves et al. analysed 36 patients with recurrent high-grade gliomas, they are treated with Gamma Knife stereotactic radiosurgery (SRS). Patients in high-risk regions of the SRS target had an improved survival rate versus those with MRS high-risk regions outside the SRS target [39]. Croteau et al. studied 31 patients whose high-grade gliomas were resected after conventional MRI and MRS. MRS can accurately define the tumor boundary using histopathologic correlation [40]. Moreover, Pirzkall et al. presented a pre-treatment analysis of 34 patients with high-grade gliomas [41]. High-risk regions defined by MRS were significantly smaller than regions in conventional T2-wighted imaging. With MRS, more areas of normal brain can be excluded from the tumour which can reduce the side effects [41].

Figure 2 illustrates the procedure of Gamma Knife SRS incorporated with MRS functional imaging. Einstein et al. conducted the first prospective Phase II trial using MRS-targeted SRS for patients treated with GBM. In this work, the value of Cho/NAA> 2 was used as the area that is required to boost doses [42]. However, the value of the Cho/NAA has no consensus to define the tumour area. The metabolic abnormality of Cho/NAA often exceeds the enhancement range of the lesion, and sometimes even exceeds the abnormal range of FLAIR [38,43,44]. In some cases, the tumor recurs where the Cho/NAA is normal [43]. Therefore, this technique has been discussed for several decades but not been transferred into clinical practice. Other obstacles include long acquisition, low spatial resolution, low robustness of spectral acquisition, and how to consistently obtain high-quality spectral images.

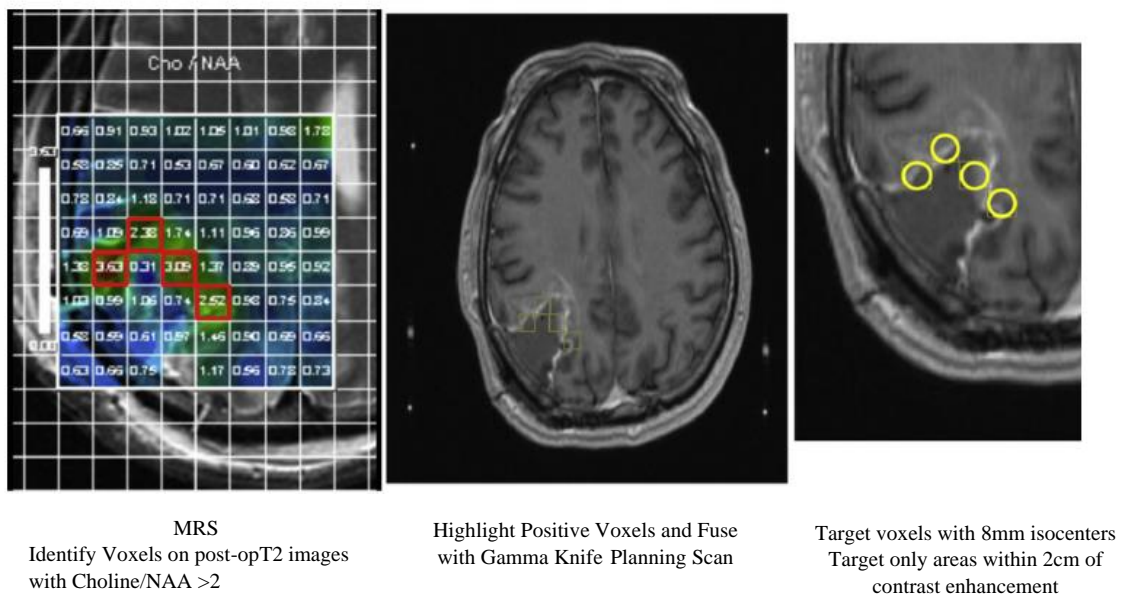

MRS
Identify Voxels on post-opT2 images
with Choline/NAA >2

Highlight Positive Voxels and Fuse
with Gamma Knife Planning Scan

Target voxels with 8mm isocenters
Target only areas within 2cm of
contrast enhancement

Figure 2. Procedure of MRS Gamma Knife SRS [42].

## 2.3 Perfusion MRI

Perfusion MRI is also a widely used method for tumour characterisation and diagnosis is [45], which includes dynamic susceptibility contrast (DSC) and dynamic contrast enhanced (DCE) MRI. When using Perfusion MRI, patients will get injection of gadolinium-based agent during continuous image acquisition. Gadolinium contrast agents can decrease the T1 relaxation time. Therefore, the distribution within the patient can be studied by continuously acquiring T1-weighted images.

Particularly, for brain tumor, modelling the distribution of contrast agent in tissue allows us to quantify vascular leakage, cerebral blood volume (CBV), mean transit time and cerebral blood flow (CBF) [46, 47, 51]. With tumor growing, new blood vessels will appear, therefore CBV, CBF and vascular leakage will be increased. In other words, CBV and CBF can be used as prognostic biomarkers to predict OS and progression-free survival PFS for tumours [46-50]. Law et al. suggested that the mean relative CBV > 1.75 can be used as the threshold of low progression time for low grade and high grade GBM [46]. Another biomarker Gd-diethylenetriamine pentaacetic acid (Ktrans) can be used to quantify the leakage of vascular thus predict OS [52]. The use of the average or median of CBF and CBV in the entire tumor volume has certain limitations for GBM as it is a highly heterogeneous tumor, which reduces the sensitivity of indicators for evaluating efficacy. For solving this problem, several methods have been proposed, such as dividing the entire tumor volume into several different sub-volumes according to different CBV values or vascular leakage [47,48], and comparing the CBV before and during RT. Prior to the use of elevated CBV for defining enhancement target in GBM, how much elevated CBV in GBM relates to tumors is a key factor. This makes sense since studies tumors can exist where CBV is not elevated [53].

Even though DSC and DCE MRI can be used to estimate CBV, there are serval limitations of perfusion MRI. Similar with DW-MRI, DSC MRI has serious geometric distortion and signal loss because it is obtained by EPI. Moreover, CBV can be achieved from T1-weighted DCE MRI [54] but determining reliable arterial input function is still non-trivial. Use of a large sagittal view is able to determine the arterial input function of aorta, thereby reducing uncertainties. To incorporate CBV calculations into the RT workflow, a validated software is also necessitated. Last but not least, consider the fact that longer scanning time for every picture empowers large spatial resolution, while a high temporal resolution restricting the spatial resolution [55-57]. Therefore, temporal and spatial resolution needs to be balanced.

## 2.4 DTI-MRI

Diffusion tensor imaging (DTI)-MRI is one of the most popular technique to detect brain tumour and implement radiotherapy. Kelly et al. [58] and Price et al. [59] have shown that tumour cells preferably transmit along the white matter tracts (WMT) and have decreased infiltration in gray matter. Krishnan et al. investigated another retrospective article showed the process from the original tumor to the recurrence using DTI-MRI for glioma patients. The results also confirmed the WMTs as a route of tumor spread [61]. These findings demonstrated that GBM grows along with WMTs. Therefore, WMT could be used as a biomarker to simulate tumor growth, and DTI-MRI can be used to define biological CTV, which have been proved by serval retrospective studies [60-64]. Moreover, DTI-MRI incorporated with mathematical model could define the RT target areas and evaluates the coverage rate of recurrence in patients with biological clinical target volume (CTV) derived from DTI.

Trip et al. implemented the first phase 0 study to deline biological CTVs using the DTI in post-operative chemo-/radiotherapy for GBM [64]. The results showed the biological CTVs derived by DTI-MRI performed worse for the central recurrences, but better for non-central and satellite recurrences. However, this study lack of enough patient samples to conclude fair results. In addition, they did not use the deformable registration, and measured the Hausdorff distance (HD) geometrically rather than anatomically. Nonetheless, due to the time-dependent migration of tumor cells, the effect of adopting a new target definition can only be truly evaluated in interventional studies.

## 2.5 PET-CT

Positron emission tomography (PET)-CT has been a valuable technique for RT in staging [66] and accurate target-volume delineation [67,68]. When patients using PET-CT, radioactive substance emitting positron must be injected. Thereafter, the scanner detects the emitted photon pairs (511 keV) and quantifies their distribution throughout the patient after signal correction and normalization steps. A variety of PET radiotracers are available in vivo biologic and molecular processes. Up to now, there are only five oncologic indications - [$^{18}$F]-fluoro deoxy-D-glucose (FDG), Na[$^{18}$F], $^{18}$fluciclovine, [$^{11}$C]-choline, and [$^{68}$Ga]-DOTA-octreotate (DOTATATE) - approved by the U.S. Food and Drug Administration. However, many other candidates are being evaluated for clinical treatment.

The details of FDG-PET are discussed in the following, since it is the most widely used PET radiotracer in the clinic. The general findings for FDG-PET still hold for other approaches. FDG-PET depends on the relation between glucose metabolism and the upregulation of glucose transporters in cancer cells, and has played a significant role for patient staging, selection, and RT target delineation [69-76]. For example, mistreatment [$^{18}$F]-FDG PET scan serves as a biomarker for adaptive dose painting. Kong et al. recently performed a phase II RT trial based on PET-CT for patients with NSCLC using interim [$^{18}$F]-FDG PET to identify regions of poorly responding disease [77]. An improved 2-year control rate was achieved with a factor of 82% versus the 69% in the Radiation Therapy Oncology Group (RTOG) 1106. Given the aforementioned results, FDG-PET is now the basis for RTOG 1106. Generally, the higher the uptake of FDG is, the more dose is needed, with a 10-30% increase to achieve the same control probability as the low FDG area [78].

Vogelius et al. derived dose–response functions for different structures that were distinguished by pre-treatment FDG-PET CT [79]. Such dose response functions are used in the dose optimization process. Versus traditional treatments, they can maximize the patient's TCP under the constraint of a constant average dose. This approach is actually a dose painting by contour (DPBC) technique, rather than a dose painting by number (DPBN) technique, i.e. without the consideration of dose–volume effect. Base on this study, Vogelius et al. implemented dose painting by numbers (DPBN) [80]. The results showed that TCP values were increased compared to uniformly delivered dose. To avoid the uncertainties existing in the relationship between SUV and TCP, Vogelius et al. focused on the heterogeneity of SUV rather than absolute SUV. This could offer us a method when the threshold of SUV is uncertain.

There is a software called Auto-PERCIST developed by Johns Hopkins University and Washington University in St Louis. This software could integrate FDG-SUV into the RT planning system [81]. This milestone will highly assist the clinicians and researchers using multi factors about tissue characterization to make decision with respect to RT dose escalation and volumes boost dose.

# III. Dose Painting

In recent decades, the advancement of RT has mainly benefited from advanced imaging technologies such as CT and MRI. Traditional tumor targets such as gross tumour volume (GTV) and clinical target volume (CTV) were defined in ESTRO-ACROP [82]. One of the "central laws" of RT is to give uniform radiation doses to the target area, but in recent years this law has been greatly challenged [88]. Many factors affect the sensitivity of radiotherapy in the tumor, such as hypoxic area, cell proliferation rate, tumor cell density and intratumor blood perfusion, etc. [89]. These factors change dynamically with the time and space during the treatment process [82]. In view of the large heterogeneity of biological characteristics in tumors, when using uniform radiation dose for heterogeneous tumors, the local recurrence will occur after radiotherapy. Therefore, it is possible to improve the local control by performing supplemental irradiation for the biological volumes that are relatively insensitive to treatment, in the tumor. Given the aforementioned discussion in Section 2, modern biological imaging techniques, such as PET-CT, and multi-parameter MRI, can facilitate the development of dose painting [83]. Dose painting is a new RT approach that produces optimized non-uniform dose distribution by using functional imaging for tumour control [83]. Dose painting can be employed for three-dimensional (3D) radiobiological analysis, thereby investigating relations among relevant parameters in RT, the inherent potential to trace the real target volume, and therapeutic dose to control the disease. At present, the supplementary exposure of dose painting to biological target areas is mostly implemented through PET-CT imaging technology.

In order to deliver a relatively higher proportion of dose to a more resistant part of tumor, dose escalation and dose redistribution have been proposed recently. There are two main strategies of dose painting, dose painting by contours (DPBC) based on threshold of biomarkers and dose painting by numbers (DPBN) based on voxels. In DPBC, a tumour's sub-volumes are heterogeneous in the functional images needed to be treated in a differentiated dose level [84]. In DPBN, dose prescription is delivered to each voxel of a tumour, determined by the voxel value in functional images. Such voxel-based dose distribution is usually represented in a dose-prescription map [84]. In the following, their definitions, features and state-of-the-art advances are reviewed.

## 3.1 DPBC

In 2005, Ling et al. proposed the first DPBC technique [6]. DPBC applies a dose boost by a certain threshold to a subvolume of the tumour. The regions of relatively lower and higher risk for recurrence are set fixed with the threshold from the quantitative functional imaging. There are many uncertainties and unknowns related to the imaging modalities where the prescription function should be based, and other unknown factors when translating the image into a prescription function for dose painting [85, 86]. Some major uncertainties, induced by PET imaging partial volume effect (PVE), tumour deformable image registration (DIR), and variation of the time interval between FDG injection and PET image acquisition have been investigated by Chen et al. [90].

In particular, interobserver variability in target volumes is a well-known factor for radiotherapy. For PET-CT based dose painting, detailed contouring guidelines can be referred to the PET-Plan (NCT00697333) clinical trial protocol [91]. As a part of the PET-Plan quality assurance, a contouring dummy run (DR) was performed to analyse the interobserver variability [92]. There is commercial software that provides the sub-target area that defines the replenishment volume and designs the RT plan [9]. In addition, Korreman et al. investigated dose conformity by using the RapidArc optimizer and beam delivery technique [87]. Optimisations for dose panting were performed in Eclipse by tuning the leaf width of multi-leaf collimator (MLC), the number of arc and collimator rotation, and by positioning uncertainties are considered for robust analysis, demonstrating low positional error [87].

Generally, a common-used treatment planning approach for DPBC is simultaneous integrated boost (SIB) technique [93]. Troost et al. proposed a DPBC-based dose escalation method to increase dose at a relatively small subvolume of the tumour, where dose threshold of the surrounding normal tissues is considered. Patients treated by this approach are expected to achieve better dose tolerance [94]. Therefore, the strategy that a homogeneous boost dose is assigned to the subvolume has been fulfilled in many clinical trials [95].

PET-guided DPBC was used to assess the feasibility of intensity-modulated radiotherapy (IMRT), where the maximum tolerated dose in head and neck cancer can be determined [96]. The same purpose of increasing tolerated dose is also explored in non–small-cell lung cancer (NSCLC) treatments. Fleckenstein et al. proposed a source-to-background contouring algorithm for FDG-PET in the process of RT planning [97]. There have been other advantages when using DPBC for RT planning. For example, Kong et al. demonstrated that adapting RT by boosting dose via DPBC to the FDG avid region improves 2-year local-regional tumour control and overall survival rate [98].

## 3.2 DPBN

The term "dose painting by numbers" has evolved to distinguish the prescription dose on a voxel-by-voxel level [99]. DPBN is a method to increase the additional dose gradually, adjusted by the local voxel intensities. Mathematical models are usually used to identify the relationship between the voxel values of the functional imaging and the risk of local recurrence [100]. In particular, Chen et al. [90] analysed how the uncertainties in quantitative FDG-PET CT imaging impact intratumorally dose–response quantification, such as ones cause by PVE and tumour DIR [90]. The negative effect arising from some of these uncertainties could be optimised by DPBN techniques [90].

There has been various technical feasibility and robustness of DPBN published recently [101–103]. Dose prescription with steep gradients can be delivered by numerous mini-subvolumes via a conventional linear accelerator [104]. Rickhey et al. used the DPBN approach in brain tumours with [$^{18}$F]-FET-PET, and satisfied treatment results were achieved with high accuracy [105]. Moreover, [$^{18}$F]-FDG-PET-guided DPBN was proved to be feasible in phase I clinical trial by Berwouts et al. in head and neck RT [106]. Recently, Grönlund et al. investigated the spatial relation between retrospectively observed recurrence volumes and pre-treatment SUV from FDG-PET [100]. Given the aforementioned findings, SUV driven dose–response functions have been presented to optimize ideal dose redistributions under the constraint of equal average dose of a tumour volume [100]. A further analysis was proposed to investigate the feasibility of DPBN to increase the TCP in a clinical scenario [107].

Some DPBN approaches have been proposed by using sub-volumes as targets [87, 100, 107] or dose maps with prescription to the voxel as objective function [108], but these methods belong to dose-volume based optimization algorithms. Jiménez-Ortega et al. presented a new optimization algorithm to implement directly constraints to voxels instead of volumes, where Linear Programming (LP) is used to carry out DPBN approximation. This method is implemented in CARMEN, a Monte Carlo (fMC) treatment planning system [84]. Since proton therapy has been reported as potentially capable of decreasing toxicity, Håkansson et al. investigated DPBN in proton RT planning by comparing proton dose distributions with delivered photon plans from a phase-I trial of FDG-PET based dose-painting [109]. Experimental results stay in line with the physical properties of the photon and proton beams, i.e. proton DPBN can be optimised with a quality comparable to photo DPBN [109].

Table 1. A review of the sate-of-the-art DPBC and DPBN techniques.

|      | Author | Year | Tumour place | Level of dose escalation | Conclusion |
|------|--------|------|--------------|--------------------------|------------|
| DPBC | Schimek-Jasch et al. [92] | 2015 | NSCLC | 60-74 Gy | Target volume delineation is improved. |
|      | Heukelom et al. [110] | 2013 | Head and neck | BR 77Gy, PTV outside the BR 67 Gy | 5% improvement in LRC with a power of 80% at a significance level of 0.05. |
|      | Kong et al. [98] | 2013 | NSCLC | 84 Gy (median) | 2-year rate of in-field LC and overall LC were 84%and 68%, the rate of OS was 51%. |
|      | Fleckenstein et al. [97] | 2011 | NSCLC | 66.6 to 73.8 Gy | Median survival time was 19.3 months |
|      | van Elmpt et al. [111] |  | NSCLC | BR 86.9±14.9 Gy | N/A |
|      | Korreman et al. [C] | 2010 | NSCLC | 90 Gy (mean) | Good conformity was obtained using MLC leaf width 2.5 mm, two arcs, and |
| | | | | | collimators 45/315 degrees, and robustness to positional error was low. |
| | Madani et al. [96] | 2006 | Head and neck | 72.5, 77.5 Gy | Actuarial 1-year rates of LC were 85% and 87%, and 1-year rate of OS was 82% and 54% (P=0.06) |
| DPBN | Chen et al. [90] | 2020 | HNSCC | N/A | Uncertainties in quantitative FDG-PET/CT imaging feedback arising from PVE and DIR have been analysed. |
| | Håkansson et al. [109] | 2020 | Head and neck | 85.3 Gy (Maximum) | Proton dose-painting can reduce the non-target dose generally, but shoud avoid unintended hot spots of mucosal toxicity. |
| | Grönlund et al. [112,107, 100] | 2020, 2019, 2017 | Head and neck | CTVT 66 to 74.5 Gy | TCP values increased between 0.1% and 14.6% by the ideal dose redistributions for 59 patients |
| | Jiménez-Ortega et al. [84] | 2017 | NSCLC | 68 Gy (minimum) | The total planning time spent ranged from 6 to 8 h. |
| | Berwouts et al. [113] | 2013 | Head and neck | Prescription dose of GTV 70.2 Gy (median) | Disease control in 9/10 patients at a median follow-up of 13 months |
| | Madani et al. [114] | 2011 | Head and neck | 80.9 and 85.9 gy (median) | Actuarial 2-year rates of LC and freedom from distant metastasis were 95%, 93% and 68%, respectively. |
| | Meijer et al. [115] | 2011 | NSCLC | 66 Gy | DPBN can increase higher dose levels than DPBC when considering organs at risk. |

Abbreviations: BR: boost region; CTVT: primary clinical target volume; LC, local-regional control; OS: overall survival; TCP: tumour control probabilities; GTV: gross tumour volume; HNSCC: squamous cell carcinoma of head and neck; PVE: partial volume effect; DIR: deformable image registration.

## 3.3 Comparisons

DPBC mainly refers to the specific function image parameters to set the threshold for the replenishment area. Biomarkers in the high-risk area for recurrence have larger values over the defined threshold, while low-risk recurrence area corresponds to biomarkers having smaller values than the threshold. DPBN assumes that the recurrence risk of a certain pixel in the tumor area is positively correlated with the parameter intensity of its specific function image pixel, and the radiation dose of a certain pixel is directly related to its corresponding functional image pixel information. DPBN directly relies on theragnostic imaging [117]. DPBN requires a customised software package to optimize the irradiation plan, but there is no commercial software directly implement optimization of DPBN [116, 120].

We remark that DPBC and DPBN have their own advantages and disadvantages [120]. Advantages of DPBC are that sub volumes that need to boost dose can be pre-drawn before the treatment plan. Then the sub volumes can be set to add margins to supplement the geometric uncertainty, and the treatment plan can also be evaluated by conventional DVH. Disadvantages of DPBC mainly include the lack of consensus of the threshold for biomarkers. DPBN has more theoretical advantages than DPBC because it can deliver doses to voxel level. However, it cannot extend the margin of specific voxels and is more sensitive to uncertainty arising in image registration. Therefore, online image-guided treatment that can clearly show soft tissues is required. Calibration of baseline displacement of the tumor is required before each treatment.

Meijer et al. examined both DPBC and DPBN techniques for non-small cell lung cancer (NSCLC) patients' treatment. In general, the amount of DPBC dose-boosting is limited whenever the GTV boost is close to any serial risk organ. However, DPBN shows significant higher dose values to high SUV voxels and are more distant from the organs at risk, since DPBN boosts work at a voxel-by-voxel basis [115].

# IV. Challenges and Future Prospect

## 4.1 Robust Calibration for Biomarkers

Current quantitative imaging approaches have larger voxels and worse signal-to-noise-ratio (SNR) compared to other clinical imaging techniques, since quantitative imaging requires one or multiple quantitative parameters for each voxel. For dynamic contrast-enhanced MRI/CT, the acquisition of multiple images is required [45]. Therefore, target delineation and dose painting on functional images rely more on the values of parameters of biomarkers compared to other clinical imaging. As was reviewed in Section 2, the main limitation of functional imaging lies in the uncertainties related to the relationship between biomarker and their corresponding parameters. Further studies involving more robust calibration are needed to propose a more robust relationship. One alternative way is to use the heterogeneity of the biomarkers rather than the absolute value of biomarker, which can reduce the impact the uncertainties of the biomarkers [80]. For tumor types with high tumor heterogeneity, such as GBM, it makes more sense to consider the biomarker parameters of voxel level instead of the mean value of a whole tumor. Ideally, dose painting prescriptions should be based on dose response data which can be observed from multiple functional imaging methods [116].

## 4.2 Dose Painting for Adaptative RT

Most RT plans remain "unchanged". Once the plans are initialized, they are carried out until the end of treatment, along with minimal modifications on top of the original plans [121]. RT plans are usually delivered for the period of several weeks, which is a long and arduous process. In most cases, the tumour size, location and microscopic status (e.g. metabolism and hypoxia) will change, thereby resulting in long-term and even life-long consequences on patients' life quality after treatment. To address such challenges, the adaptive RT has been developed recently, where interests in functional image-based dose painting have been growing. Many studies have discussed the potential of functional images for a more personalized RT planning, however, how to incorporate such quantitative functional imaging into adaptive RT has been barely investigated. For example, since functional imaging can define easy-to-control and hard-to-control areas in the tumor before a treatment, assume that patients take the same functional imaging after a period (e.g., 2 weeks) of radiotherapy, changes in biomarkers can be measured, indicating radio sensitivity and radio resistance after serval fractions of radiotherapy. Adaptative RT can be therefore realisable.

## 4.3 Alleviation of Uncertainties

The field of radiation oncology has been used to address different sources of uncertainties during patient planning, including range and setup uncertainties [122], organ movements [123] and clinical target volume (CTV) definition [124]. Consider that the implementation of dose painting depends on the quality of data obtained via biomarkers. The interpretation of such data inevitably introduces some uncertainties. Therefore, it is foreseen that potential mathematical tools to solve uncertainties, e.g. partially observable Markov decision processes (POMDP), imperfect state information (ISI) and adjustable robust optimization (ARO) might be necessitated.

# V. Conclusion

In this article, we reviewed the state-of-the-art functional imaging techniques which facilitates the development of dose painting. Dose painting by contour and dose painting by numbers are discussed in detail, respectively, along with a summary of their advantages and disadvantages. Finally, we conclude some existing challenges and provide corresponding possible research directions, where we hope this work can lead to fruitful research avenues in this thriving field.

# Acknowledgement

The authors are grateful for the support of the China Scholarship Council. They also wish to thank Dr. He Li for his helpful suggestions.